\newif\ifsubmode
\submodefalse

\newif\ifprintfig
\printfigtrue

\newif\ifemulate
\emulatetrue
\ifsubmode
\documentclass[12pt,preprint]{aastex}
\received{}
\accepted{}
\journalid{}
\articleid{}
\else
   \documentclass{emulateapj}
   \submitted{{\it Accepted for publication in ApJ}}
\fi
\usepackage{multirow,color,natbib,wrapfig,ulem}
\def\lsim{\mathrel{\raise.3ex\hbox{$<$\kern-.75em\lower1ex\hbox{$\sim$}}}}
\def\gsim{\mathrel{\raise.3ex\hbox{$>$\kern-.75em\lower1ex\hbox{$\sim$}}}}
\newcommand{\Msun}{\mbox{\,$M_{\odot}$}}

\newcommand{\kms}{\ifmmode \,{\rm km~s}^{-1} \else \,km~s$^{-1}$\fi}
\newcommand{\Rvir}{\ifmmode R_{\rm vir} \else $R_{\rm vir}$\fi}
\newcommand{\vmax}{\ifmmode v_{\rm max} \else $v_{\rm max}$\fi}
\newcommand{\vpeak}{\ifmmode v_{\rm peak} \else $v_{\rm peak}$\fi}
\newcommand{\vinf}{\ifmmode v_{\rm infall} \else $v_{\rm infall}$\fi}
\newcommand{\vone}{\ifmmode v_{\rm 1kpc} \else $v_{\rm 1kpc}$\fi}

\def\spose#1{\hbox to 0pt{#1\hss}}
\def\simlt{\mathrel{\spose{\lower 3pt\hbox{$\mathchar"218$}}
     \raise 2.0pt\hbox{$\mathchar"13C$}}}
\def\simgt{\mathrel{\spose{\lower 3pt\hbox{$\mathchar"218$}}
     \raise 2.0pt\hbox{$\mathchar"13E$}}}

\shorttitle{No Massive Missing Satellites}
\shortauthors{Brooks et al.}

\begin{document} 

\title{A Baryonic Solution to the Missing Satellites Problem}

\author{Alyson\ M.\ Brooks\altaffilmark{1}, Michael\ Kuhlen\altaffilmark{2}, Adi\ Zolotov\altaffilmark{3}, Dan\ Hooper\altaffilmark{4}}

\altaffiltext{1}{Department of Astronomy, University of Wisconsin-Madison, 475 N. Charter St., Madison, WI 53706; 
  abrooks@astro.wisc.edu}
\altaffiltext{2}{Theoretical Astrophysics Center, University of California, Berkeley, CA, 94720;
  mqk@astro.berkeley.edu}
\altaffiltext{3}{Racah Institute of Physics, The Hebrew University,
  Jerusalem, Israel 91904; zolotov@phys.huji.ac.il}
\altaffiltext{4}{Theoretical Astrophysics, Fermi National Accelerator Laboratory, P.O. Box 500, Batavia, IL, 60510; 
  dhooper@fnal.gov}

\date{\today}

\begin{abstract}
It has been demonstrated that the inclusion of baryonic physics can alter the 
dark matter densities in the centers of low-mass galaxies, making the central 
dark matter slope more shallow than predicted in pure cold dark matter simulations.  
This flattening of the dark matter profile can occur in the most luminous 
subhalos around Milky Way-mass galaxies.  \citet{zolotov} have suggested a 
correction to be applied to the central masses of dark matter-only satellites 
in order to mimic the affect of (1) the flattening of the dark matter cusp due 
to supernova feedback in luminous satellites, and (2) enhanced tidal stripping 
due to the presence of a baryonic disk.  In this paper, we apply this correction 
to the $z=$0 subhalo masses from the high resolution, dark matter-only 
Via Lactea II (VL2) simulation, and find that the number of massive subhalos is 
dramatically reduced.  After adopting a stellar mass to halo mass relationship 
for the VL2 halos, and identifying subhalos that are (1) likely to be destroyed 
by stripping and (2) likely to have star formation suppressed by photo-heating, 
we find that the number of massive, luminous satellites around a Milky Way-mass 
galaxy is in agreement with the number of observed satellites around the 
Milky Way or M31.  We conclude that baryonic processes have the potential to 
solve the missing satellites problem.  
\end{abstract}

\section{Introduction}

The cosmological paradigm based on cold dark matter (CDM) and dark energy 
($\Lambda$) has been extremely successful in describing the observed evolution 
and large scale structure of our Universe. At small scales, however, a number 
of observations seem to be at odds with the predictions 
of $\Lambda$CDM cosmology. In particular, over a decade ago it was pointed out 
by \cite{Moore1999} and \cite{Klypin1999} that the number of high mass subhalos 
predicted by high resolution CDM simulations exceeds the observed number of 
luminous satellites of the Milky Way (MW) by at least an order of magnitude. This has 
become known as the ``missing satellites problem'' (MSP).  Although this problem has 
been mitigated to a degree by the discovery of a number of additional faint 
satellite galaxies~\citep{Willman2005,SatData9,SatData11,SatData13,SatKinematics,
SatData14,SatData15,SatData16,SatData18,SatData19}, there remains a considerable 
discrepancy between the number of observed MW satellites and the number 
predicted in CDM simulations. 

Efforts to resolve this issue have fallen into two broad categories. First, there 
are proposals in which the star formation rate in satellite galaxies is suppressed, 
leading to large numbers of low mass subhalos which are simply unobservable. 
Possible means for such suppression include photoevaporation resulting from 
ionizing radiation, e.g., at reionization \citep{Quinn1996, Thoul1996, Navarro1997, 
Klypin1999, Barkana1999, Gnedin2000, Hoeft2006, Madau2008, Alvarez2009} or 
lower $z$ due to blazars \citep{Pfrommer2012}, or due to cosmic ray heating 
\citep{Wadepuhl2011}.  Photoionization is 
expected to suppress star formation in halos below $\sim$10$^9$ M$_{\odot}$ 
\citep{Okamoto2008}.  In more massive halos where gas is retained and star 
formation can begin, further suppression is expected from supernova feedback 
\citep{Dekel1986, Benson2002, Dekel2003c, Governato2007}.  It is also important 
to consider the number of faint satellites that remain undetected due to 
observational incompleteness \citep{Willman2004, SatKinematics, Tollerud2008, 
Walsh2009, 
Koposov2009, Rashkov2012}.  Taken together, it is possible that observational 
incompleteness combined with suppression of star formation can bring the number 
of luminous satellites in line with the number of predicted satellites around 
a Milky Way-mass galaxy.  

However, while this combination of effects may explain the faint or
low-mass regime of the MSP, \citet{toobig} recently pointed out that
it still fails at the massive end. Dubbed the ``too big to fail'' (TBTF) 
problem, this aspect of the MSP arises because the most massive subhalos in 
ultra-high resolution dark matter-only simulations of MW-analog
galaxies \citep{Diemand2008,Springel2008} are too dense to host the observed 
satellites of the MW \citep[see also][]{Wolf2012, Hayashi2012}. 
The simulations always contain a population of subhalos (6-22, varying within 
the errors of the MW's measured mass) that are more massive than any of the 
dwarf spheroidals observed in the MW \citep{Boylan-kolchin2012}.
In other words, while the abundance of lower luminosity satellites may be 
made consistent with CDM predictions, the MSP remains a puzzle because 
simulations predict too many \textit{massive} satellites.

A second class of possible solutions to the MSP considers departures from
the standard assumptions of cold and collisionless dark matter. Whereas typical 
particle dark matter models predict that WIMPs will form halos with masses as 
small as $10^{-6}\,M_{\odot}$ or so (depending on the temperature at which the 
WIMPs undergo kinetic decoupling with the cosmic neutrino background), the
formation of small scale structure can be strongly suppressed if the dark
matter particles are not entirely cold~\citep{Colin2000,MaccioFontanot2010,
Lovell2012}.  Dark matter in the form of sterile neutrinos with masses of
$\sim$1-10 keV have received attention within this context~\citep{sterile},
although many other warm dark matter (WDM) scenarios could potentially
accomodate a similar suppression of small scale
power~\citep{gravitino,mev}. WDM primarily addresses the low-mass end of
the MSP by suppressing the formation of small scale structure. It can also
help at the high mass end, however, because the delayed halo collapse times
in a WDM cosmology result in lower concentrations and hence reduced central
densities \citep{Lovell2012}.

Self-interacting dark matter (SIDM)
models \citep{Carlson1992,Spergel2000,Loeb2011} provide another mechanism
to achieve the same effect. In this case, the interactions prevent the
formation of the steep central density cusps that are the hallmark of CDM
halos. Density profiles in SIDM instead exhibit a central
core \citep{Vogelsberger2012,Rocha2012,Zavala2012}, which helps to address the MSP in
two ways: cored halos are more susceptible to tidal disruption, potentially
removing many of the excess low mass halos; secondly, in the surviving
halos, the core reduces the central densities, alleviating the TBTF
problem \citep{Vogelsberger2012}. Furthermore, there is considerable
observational evidence for the existence of dark matter cores in the
centers of low surface brightness
galaxies \citep{KuzioDeNaray2008,deBlok2010,Oh2011} and in at least two
MW dwarf spheroidal satellites \citep{Walker2011}.

In this work we consider a new type of solution to the MSP in which the 
shape of satellite dark matter profiles are altered, but as a
result of baryonic physics, rather than through modifications in the
particle physics sector.  This model incorporates suppression of star
formation from photoionization and supernova feedback, but additionally
considers the tidal effects due to the presence of a baryonic disk, which
is not found in dark matter-only simulations.  The loss of gas in satellite
halos through tidal stripping has been proposed as a means to limit star
formation~\citep{Redef_MSP}.  However, tidal stripping in a cuspy halo by
itself does not reduce the central densities of the most massive subhalos
enough to bring them into agreement with the observed kinematics of the
MW satellites \citep{Read2006b,Boylan-kolchin2012}. An additional 
modification to the central densities of the most massive satellites 
appears to be required.

It has become broadly accepted in recent years that baryonic 
processes can alter the distributions of dark matter within halos, potentially 
steepening \citep{Blumenthal1986, Gnedin2002, Gnedin2011} or flattening 
\citep{Navarro1996, Read2005, Mashchenko2006, Mashchenko2008, Governato2010, 
Pasetto2010, deSouza2011, Cloet2012, Maccio2012, Pontzen2012, Governato2012, 
Teyssier2012} 
the profile depending on the mass of the halo in question, and on the strength 
of various feedback mechanisms.  Recently, the effects of baryons on the dark 
matter halo profiles of luminous satellite galaxies was studied by~\cite{zolotov}. 
These authors found that satellites with $M_{\rm star} \gsim 10^7 \, M_{\odot}$ 
tend to develop ``cored'' density distributions through supernova feedback prior 
to infall \citep[here, cored refers to any inner density slope, $\gamma$, 
shallower than -1, because $\gamma$ becomes flatter with increasing stellar 
mass, see][]{Governato2012}. The flatter density profiles
make them more vulnerable to tidal effects of the baryonic disk 
\citep[see also][] {Taylor2001, Read2006b, Choi2009, Wetzel2010, Romano2010, 
D'Onghia2010, penarrubia}.  \cite{brookszolotov} showed that the presence of 
the disk increases tidal stripping for galaxies of all masses 
compared to the dark matter-only case, but most strongly reduces the central 
densities of cored satellites \citep[see also][]{Stoehr2002, Hayashi2003, 
penarrubia}, which are the most luminous. In other words, baryonic physics 
significantly reduces the observable mass of satellites. \cite{brookszolotov} 
demonstrated that these combined baryonic effects produce a $z=0$ satellite 
distribution with luminosities and kinematics comparable to the MW and M31.

In this work, we apply the results of \citet[][hereafter Z12]{zolotov} to 
the subhalo populations found in the DM-only Via Lactea II simulation.  We find that after 
taking into account baryonic effects, the number of surviving massive subhalos is 
strongly reduced.  Adopting a stellar mass to halo mass relation, we show that the 
number of luminous satellites is also compatible with observations.  This suggests that a proper 
accounting of baryonic effects reduces the predicted number of massive, luminous 
subhalos that should survive in CDM, bringing the observations and theory into 
agreement without the need of invoking warm dark matter or dark matter with 
exotic properties or interactions.


\section{Satellite Masses}

We make use of the publicly available subhalo
catalogs\footnote{http://www.ucolick.org/$\sim$diemand/vl/} from the Via
Lactea II (VL2) DM-only simulation \citep{Diemand2008}, supplemented by
additional information extracted from the simulation outputs, consisting of
(1) mass enclosed within 1 kpc at $z=0$, (2) infall times and the maximum
value in their rotation curves, \vmax, at infall, and (3) their full
orbital information \citep{Kuhlen2012b}, including number and distance of
apo- and pericenter passages.  The VL2 halo was run using a WMAP year 3
cosmology \citep{Spergel2007}.  The halo was selected to have no low $z$
major mergers, thought to be similar to the merging history of the MW.  The
$z=0$ halo has roughly 450 million particles with masses of 4100
M$_{\odot}$ within its virial\footnote{Here we adopt a virial overdensity
of $\rho/\rho_0 = 200$.} radius of 402 kpc at $z=0$, yielding a halo mass
of 1.9$\times$10$^{12}$ M$_{\odot}$.

Starting from all subhalos within \Rvir\ at $z=0$, we restrict our study to
only those subhalos with \vpeak\ (the largest value of \vmax\ over the
subhalo's entire history) $> 13.3 \kms$.  This \vpeak\ value was chosen to 
correspond to 10$^5$ M$_{\odot}$ in stellar mass, or a $V$-band absolute 
magnitude of $-7$ (see Section 3.3).  This allows our sample to cover the full 
range of luminosities of the classical dwarf satellites.  Selecting 
for $\vpeak > 13.3 \kms$ yields a sample of 410 subhalos. It is commonly assumed that 
all halos with $\vpeak < 20 \kms$ are strongly affected by UV heating, unbinding 
their gas \citep[e.g.,][]{Okamoto2008}.  This process is expected to either 
leave them entirely devoid of stars (``dark''), or very faint if they were able 
to form a few stars prior to reionization \citep[e.g.,][]{Bovill2009, 
Salvadori2009, Li2010, Wolf2010}, and prone to be missed by current surveys.  
In Section 3.2 we consider the impact of UV heating on the VL2 
subhalos, and show that, indeed, most of these low mass subhalos remain dark.  
However, a few of the satellites in this mass range should be capable of forming 
stars. 

The \vmax-function of our final subhalo sample (i.e., those halos that
survive to $z=0$ and had $\vpeak > 13.3 \kms$) is shown in Figure 1, for
various times.  The solid line shows the cumulative \vpeak-function for the
halos in our sample, while the short-dashed curve shows \vmax\ at infall.
We note that \vpeak\ and \vmax\ at infall, \vinf, can be slightly different 
for these subhalos.  The majority of the VL2 subhalos in our sample
reach \vpeak\ at high $z$ and then grow very little, with an overall slight
decrease ($<$ 10\%) in \vmax.  A few subhalos undergo a substantial
reduction in \vmax\ between high $z$ and infall, presumably due to encounters
with other halos that strip their mass. 
The resulting velocity function at $z=0$ is shown in
Figure 1 as the dotted line.  The circular velocity at 1 kpc, \vone, is
shown for comparison (long-dashed line).  In the following, we adopt \vone\
to examine the effect of baryons on the central mass distribution of 
satellite galaxies.

\begin{figure}
\includegraphics[angle=0,width=0.5\textwidth]{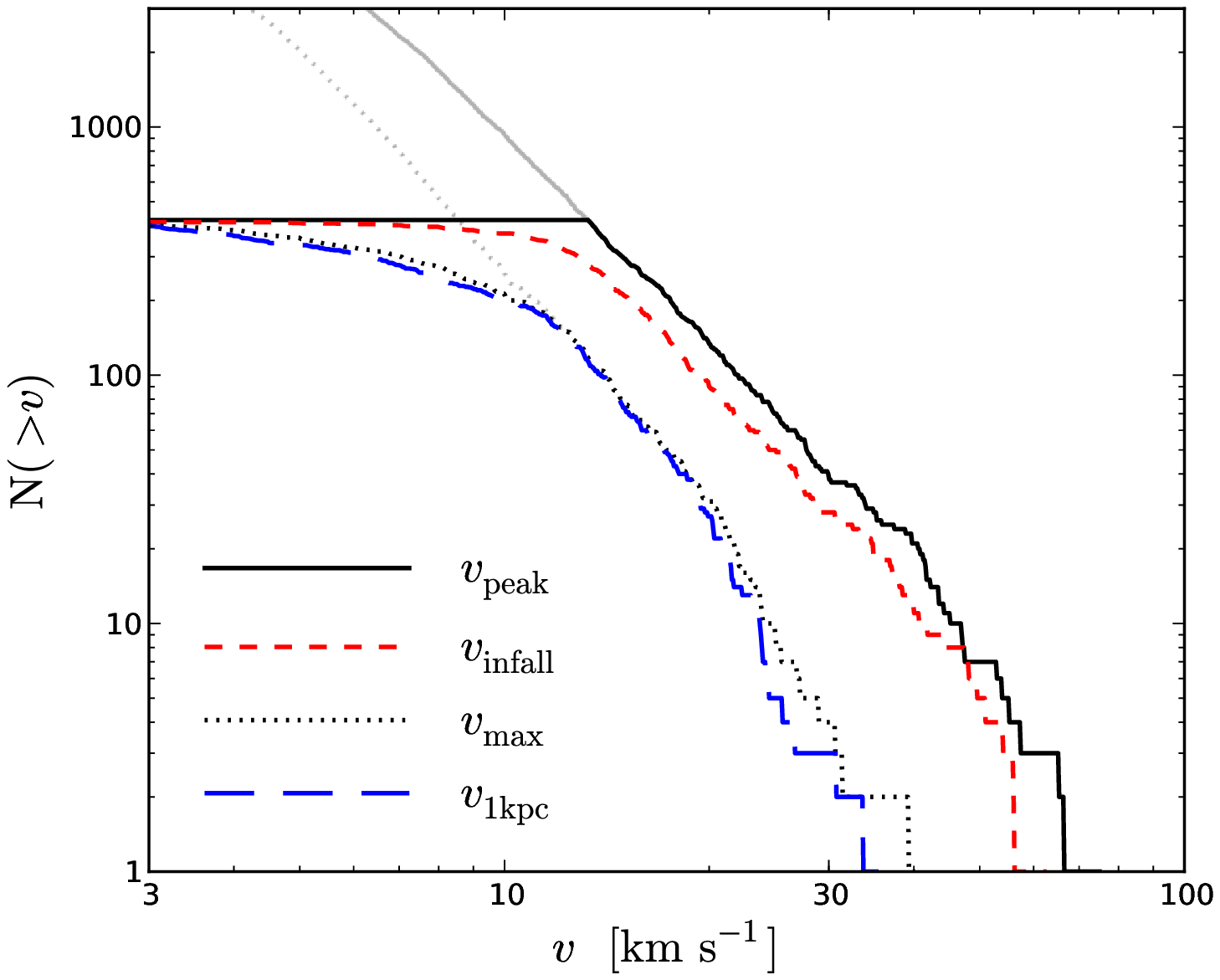}
\caption{Cumulative velocity functions for VL2 halos that survive to $z=0$. 
The \vpeak-function is plotted with a solid black line for $\vpeak > 13.3 \kms$. 
The corresponding evolved \vmax-functions are also shown at infall (red, short 
dashed line) and at $z=0$ (dotted line). For comparison, we also plot \vone\ 
at $z=0$ (blue, long dashed line).  The \vpeak\ and \vmax-functions found in 
the full VL2 catalog (no cut on \vpeak) are shown as continuations in light 
grey.}
\label{vcfunc}
\end{figure}

Z12 proposed a correction to the \vone\ values of $z=0$ DM-only subhalos
to account for the missing baryonic physics that lowers the central 
masses of luminous subhalos,  
\begin{equation}
\Delta(\vone) = 0.2v_{\rm infall}-0.26 \, {\rm \kms}.
\end{equation}
For subhalos with $\vinf < 30 \kms$, the Z12 correction is designed 
to account for a reduction in central mass due to (1) loss of gas, either due 
to UV heating before infall or stripping after infall, and (2) the tidal effect 
of the baryonic disk (which does not exist in DM-only runs).  These processes 
should act on a subhalo even if it is too low mass to retain gas at
reionization, or massive enough to be luminous.  These processes should also 
occur irrespective of whether the subhalo has a cuspy or a cored DM density 
profile.  Yet neither are typically accounted for in DM-only simulations.  In 
subhalos with $\vinf > $ 30 \kms, Z12 found that enough star formation takes 
place to significantly flatten the DM density profiles prior to infall.  Hence, 
at $\vinf > $ 30 \kms, the Z12 correction accounts for (1) and (2) as in the 
lower mass case, but also for an additional reduction in the central masses 
of the satellites due to supernova feedback and the enhanced tidal stripping 
that occurs as the central density profiles of massive satellites become more 
shallow.

In Z12, equation 1 was derived using halos 
with $20 \kms < \vinf < 50 \kms$.  Hence, it has not been tested down to the 
lower $\vinf$ of our lowest mass VL2 subhalos.  However, we will apply it to 
all subhalos with $\vinf < 50 \kms$, and in Section 3 we show 
that in all cases where the correction yields unphysical (negative) results, 
the unphysical halos lose more than 97\% of their mass at infall, and should 
be considered completely destroyed in the tidal presence of a baryonic disk.   
At $\vinf >  50 \kms$, subhalos 
are Magellanic-like and gas-rich at accretion, possibly including an additional 
effect of adiabatic contraction that is not accounted for in the correction.  
There are 5 massive VL2 subhalos with $\vinf > 50 \kms$, for which the Z12 
correction is not applied. 

Most of the MW and M31 dwarf spheroidal galaxies (dSphs) have 
central $v_c$ values $<  20 \kms$ \citep[measured at the half light radii, 
which are all $\lesssim$ 1 kpc,][]{McConnachie2012}.  At $z=0$, the VL2 
host halo contains 28 satellites with $\vone > 20 \kms$, grossly inconsistent 
with the observational results.  After applying the correction of Z12, there 
are only 5 satellites with $\vone > 20 \kms$. 
This is a first indication that baryonic physics, at least as
implemented in the simulations of Z12, appears to be quite capable of
reducing the mismatch between the central densities of the most massive
subhalos in DM-only simulations and the observed kinematics in the
classical MW dwarf satellite galaxies.

\section{Luminous Satellites at Redshift 0}

It is clear that the baryonic correction proposed by Z12 dramatically reduces 
the number of massive halos expected in a DM-only MW-mass run (e.g., from 28 
to 5 for satellites with $\vone > 20 \kms$ in VL2).  However, it is 
less clear if it reduces the number of luminous subhalos.  That is, does the 
correction take 20 Fornax-like satellites that initially have $\vone > 20
\kms$ at $z=0$ and simply shift them to $\vone < 20 \kms$?  If so, then there 
may be fewer {\it massive} halos, but the overall number of {\it luminous} 
satellites would still be much larger than observed in the MW or M31. In this 
section we explore whether both the mass and luminosity of Milky Way-mass 
galaxies can be reproduced by accounting for baryonic effects.

\subsection{Destruction Rates}
We must first consider whether the VL2 halos should have all survived 
to $z=0$, or whether it is likely that baryonic physics would have led to 
their destruction.  \citet{penarrubia} examined the effect of a baryonic 
disk on both cored and cuspy subhalos.  They showed that tides from the disk 
are dominant in the center of cored subhalos, leading to much more mass loss 
than DM-only runs predict.  Even cuspy subhalos will undergo more mass 
loss when a baryonic disk is present if they are on highly elliptical orbits. 
Importantly, cuspy halos tend to survive even after substantial (99.99\%) mass 
loss, while cored halos can be completely disrupted.  This implies that there 
could be a large number of satellites that survive in the DM-only VL2 run 
that would not survive in a baryonic run.

In what follows, we assume that a VL2 subhalo that has lost a certain fraction 
of its mass since infall should have been fully disrupted if baryons had been 
included, and should no longer appear as a bound luminous satellite at $z=0$.  
To set the limiting fraction, we refer to figure 2 in \citet{penarrubia}, where 
tidal effects of a disk on cored dwarf galaxies are compared to those on cuspy 
halos with no disk present.  Adopting their ``mixed'' model, in which a cored 
($\gamma = 0$) subhalo evolves within a parent halo with a cuspy ($\gamma = -1$)  
density profile, we find that cored subhalos that have pericenters $\lesssim$ 
20 kpc undergo 99.9\% mass loss, which we consider to be completely disrupted.
Without the disk presence, the results in \citet{penarrubia} show that a cuspy 
subhalo with a pericenter of 20 kpc would experience 90\% mass loss.  Hence, we 
assume that any VL2 subhalos with pericenters $<$ 20 kpc should experience 
strong disk tides that need to be accounted for.  
We conclude that if a VL2 DM-only satellite has lost more than 90\% of its mass 
after infall, and has had pericentric passages that take it within 20 kpc of the 
parent halo's center, the same satellite in a baryonic run is extremely 
likely to be fully disrupted by tides.

To be conservative, we apply this limit only to subhalos that have 
$\vinf >$ 30 \kms.  Z12 and \citet{Governato2012} demonstrated that cored DM 
density profiles exist in satellites with a stellar mass of more than 
10$^7$ M$_{\odot}$, corresponding to halos with $\vinf >$ 30 \kms, 
and consistent with the energy arguments explored in both \citet{Boylan-kolchin2012} 
and \citet{Penarrubia2012}.  While it is possible that halos at lower 
masses have cores (but at smaller radii that are below the resolution limit of 
current cosmological simulations), we assume they may remain cuspy, and 90\% mass 
loss may then not be enough to fully destroy them.  Instead, for the remainder of 
the halos with $\vinf <$ 30 \kms, we adopt the results of 
\citet{Wetzel2010} and assume that a halo must have lost 97\% of its mass since 
infall to be fully stripped, and no longer appear as a bound luminous satellite 
at $z=0$.  We verified that the majority of the halos that have lost 97\% of 
their mass have tidal radii less than 1 kpc, suggesting that their inner luminous 
regions should indeed be stripped.  Adopting 97\% is conservative, as both Z12 
and \citet{Penarrubia2008} found that halos begin to have their stars stripped 
after losing 90\% of their halo mass after infall. 

We use the difference in \vmax\ at infall and \vmax\ at $z=0$ for a 
given VL2 subhalo to estimate the amount of mass that it has lost since infall. 
Following the results of \citet{penarrubia}, 
\begin{equation}
\frac{v_{max}(z=0)}{v_{infall}} = \frac{2^{0.4}x^{0.3}}{(1+x)^{0.3}} \nonumber
\end{equation}
for a subhalo with a DM density slope, $\gamma = -1.0$, where 
$x \equiv$ mass($z=0$)/mass($z=$infall).  A slope of -1 is roughly the slope in 
the central regions for all halos formed in DM-only simulations \citep{nfw, 
Springel2008}, and we verified that the inner density slopes in the VL2 subhalos 
are also consistent with -1.  We use the above equation to find (1) all subhalos 
that lose more than 97\% of their infall mass ($x =$ 0.03) and (2) all subhalos 
with $\vinf >$ 30 \kms that lose more than 90\% of their infall mass ($x =$ 0.1) 
and have pericentric passages under 20 kpc.  We consider these two populations 
of subhalos to be ``destroyed.''

\subsection{Identifying Dark Subhalos}

As we will show below, even after considering destruction due to tidal effects 
on the VL2 subhalos, 307 satellites remain at $z=0$ from our original sample of 
410 with \vpeak\ $> 13.3$ \kms.  Roughly 100 of the surviving subhalos should be 
as bright as the Draco dSph (see next section), and thus luminous enough to have 
been detected, but 100 such luminous satellites is much larger than the number 
observed in the MW or M31 to date.  Yet many of these satellites 
are in the halo mass range that is expected to be strongly affected by UV heating, 
and some of these subhalos should therefore be inefficient at forming stars and 
will remain dark. In this section, we apply the results of \citet{Okamoto2008} to 
identify the subhalos that may remain dark.  

\citet{Okamoto2008} use hydrodynamical simulations to identify the characteristic 
halo mass, $M_{\rm char}$, that retains 50\% of the cosmic baryon fraction, 
$f_{\rm bar}$, as a function of $z$.  Their simulations adopt a uniform ionizing 
background that accounts for H and He\,I reionization at $z=9$, and He\,II 
reionization at $z=3.5$.  We have converted their $M_{\rm char}(z)$ results into 
$v_{\rm char}(z)$, adopting a WMAP3 cosmology and using an overdensity of 
200$\rho_{crit}$.  We assume that if a VL2 subhalo has a \vpeak\ value above 
$v_{\rm char}(z_{\rm peak})$ it retains enough baryons to be luminous.
Note that $v_{\rm char}(z)$ is a virial quantity, but we do not have the virial 
masses of the VL2 subhalos. Rather, we have \vmax\ values.  Most satellites have
$1.2 < \vmax$/$v_{\rm vir} < 1.8$, corresponding to a range of concentrations\footnote{A 
halo's concentration, $c$ = \Rvir/$R_{\rm s}$, relates the the virial radius to 
a scale radius, $R_{\rm s}$, where the steep, outer density profile of the 
halo transitions to the less steep, inner profile, e.g., where $\gamma \sim-2$ 
for an NFW halo.}
$10 < c < 40$, with a mean \vmax$/v_{\rm char} = 1.4$ for a halo with 
$c = 20$ \citep{Bullock2001b, Prada2012}. 

\begin{figure}
\includegraphics[angle=0,width=0.5\textwidth]{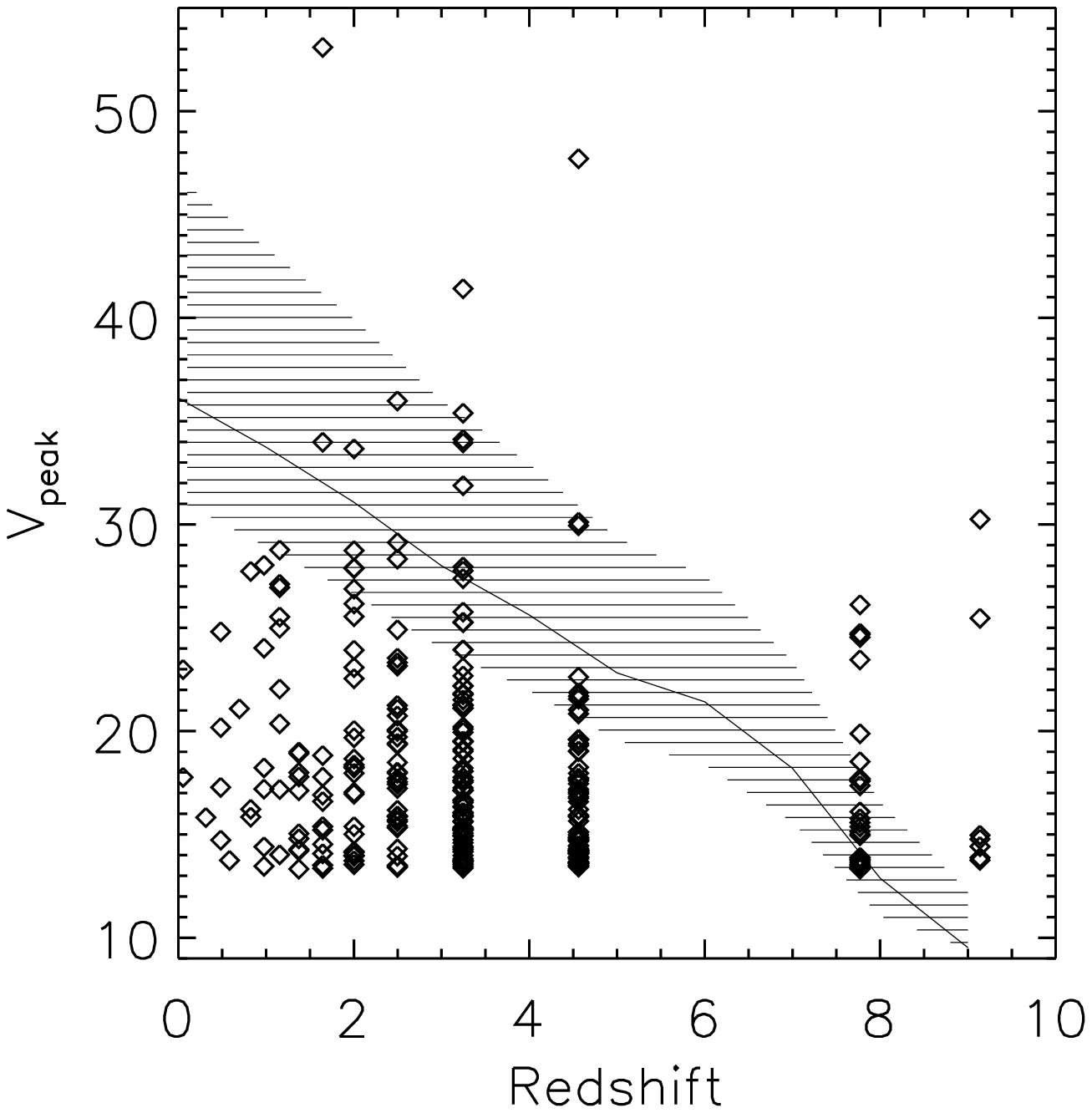}
\caption {Criteria to select dark subhalos.  \vpeak\ is shown for VL2 
subhalos that survive tidal stripping, at the redshift of \vpeak.  The 
solid line shows 1.4$v_{\rm char}$ following \citet{Okamoto2008}, 
as a function of $z$.  The shaded region shows varying 
$1.2 < \vmax$/$v_{\rm char} < 1.8$. 
For our analysis, VL2 subhalos below the solid line are considered ``dark.''} 
\label{dark}
\end{figure}

Figure 2 illustrates the affect of applying this model to the VL2 subhalo 
population.  Diamonds show the \vpeak$(z_{\rm peak})$ values of the VL2 
satellites that survive after our tidal stripping considerations.  To put the 
$v_{\rm char}(z)$ of \citet{Okamoto2008} into \vmax\ space for comparison with 
the VL2 subhalos, we have multiplied $v_{\rm char}(z)$ by 1.4 to derive the 
solid line in Figure 2 (i.e., we assume \vmax$/v_{\rm char} = 1.4$, a typical
value for a subhalo).  The shaded region surrounding the solid line shows the 
full range of $1.2 < \vmax$/$v_{\rm char} < 1.8$.  

Clearly, the number of subhalos above $v_{\rm char}$ depends sensitively on 
the concentrations (and hence \vmax$/v_{\rm char}$) of individual subhalos.  
Adopting \vmax$/v_{\rm char} = 1.4$ (1.2) leads to 40 (70) surviving, luminous  
satellites at $z=0$.  Certainly scatter in concentration is expected amongst 
the individual VL2 halos, which we have not accounted for.  Further, 
concentrations at these halo masses are expected to decrease with increasing
$z$ \citep{Bullock2001b, Eke2001, Zhao2003, Maccio2008, Prada2012}.  Again, 
we do not account for the change 
in concentration with $z$, but note that the trend would allow for lower 
\vpeak\ halos to lie above $v_{\rm char}$ with increasing $z$.  The overall 
determination of whether a particular halo is likely to be luminous will 
depend on the halo's formation and evolution.  For the purposes of this work,
we assume that all halos above the solid line in Figure 2 are likely to be 
luminous.

\subsection{Assigning Luminosities}

\begin{figure}
\includegraphics[angle=0,width=0.5\textwidth]{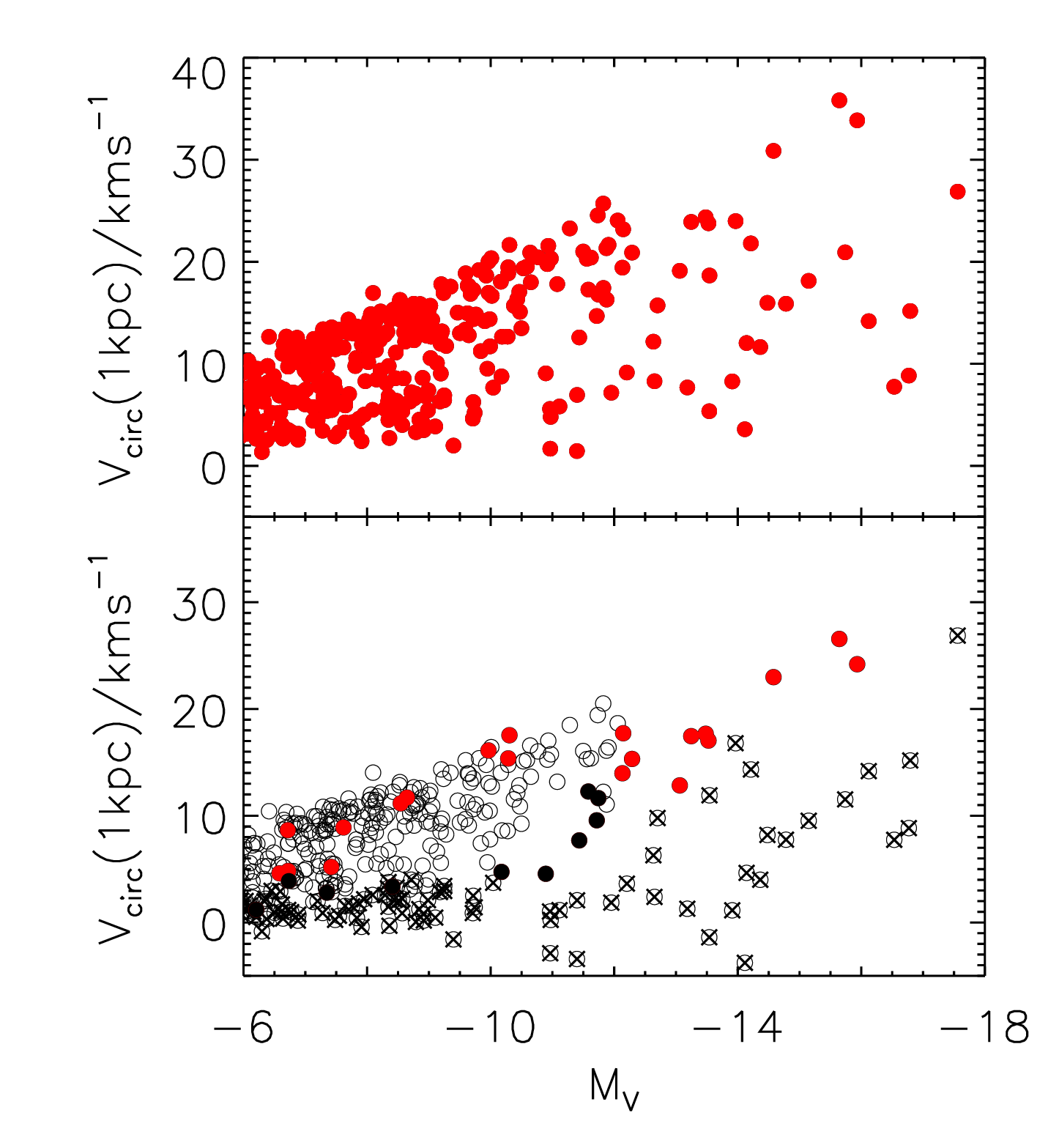}
\caption {\vone\ vs M$_{V}$ of the VL2 subhalos.  The top panel 
is the direct result from VL2 at $z=0$, while the bottom panel shows the
corrected kinematics based on Z12, along with considerations about which
subhalos are likely to be observable.  Filled red symbols are those
satellites that should be observable at $z=0$.  Subhalos that are unlikely
to survive due to the tidal effects of a baryonic disk are marked by circles 
with an x through them.  Empty symbols are subhalos that are likely to be 
dark. Filled black circles should be luminous, and do not experience enough 
stripping to satisfy our destruction criteria, but have lost enough mass 
that stars should be stripped and the luminosities should be considered 
upper limits.}
\label{vcmstar}
\end{figure}

Finally, we assess whether the surviving satellites are luminous by using
the $\vinf-{\rm M}_{\rm star}$ relation from Z12\footnote{In these simulations, 
M$_{\rm star} \propto$ M$_{vir}^2$ \citep{Governato2012}. M$_{\rm star} 
\propto$ $v_{max}^6$ because $v_{max}$ scales as M$_{vir}^{1/3}$ 
\citep[e.g.,][]{Klypin2011}. For M$_{\rm star} > $10$^6 M_{\odot}$, 
the slope of the M$_{\rm star}$-M$_{vir}$ relation adopted here lies between values 
commonly adopted in the literature \citep[e.g.,][]{Koposov2009, Kravtsov2010, 
Rashkov2012}.  However, at lower 
M$_{\rm star}$ values, abundance matching techniques used by these same authors 
suggest a much steeper relation.  We caution that equation 3 may then lead to 
overestimates of the luminosity for subhalos in Figure 3 fainter than $M_V = -10$.} 
to assign stellar masses to the VL2 subhalos, 
\begin{equation}
\frac{{\rm M}_{\rm star}}{{\rm M}_{\rm \odot}} = 0.018 \left( \frac{v_{\rm infall}}{\kms} \right)^6 .
\end{equation}
We use the tight $\log({\rm M}_{\rm star})-M_V$ for the subhalos, 
\begin{equation}
\log_{10}(\frac{{\rm M}_{\rm star}}{{\rm M}_{\rm \odot}}) = 2.37 - 0.38M_V , 
\end{equation}
in Z12 to further assign $V$-band magnitudes, $M_V$.
Z12 showed that this relation produced simulated satellite luminosity 
functions that were in good agreement with the classical dSph populations of  
the MW and M31.  Additionally, \citet{Munshi2012} have shown that the simulations 
that equations 3 and 4 are drawn from yield an excellent match to the $z=0$ 
stellar-to-halo mass relation from \citet{Moster2012}. Most
importantly, \citet{Governato2012} demonstrated that these mass-to-luminosity 
relations produced excellent agreement with field galaxies in the same 
luminosity range as the satellites we examine here.  In other words, despite 
the fact that past simulations have overproduced stars \citep{Zolotov2009, 
Guo2010, Sawala2011, Brooks2011, Leitner2012, Moster2012}, the simulations 
used to derive equations 3 and 4 are in excellent agreement with observed 
stellar-to-halo mass relations.  There is no indication that the luminosities 
predicted by these relations are too bright, though we stress that it has not 
been tested fainter than the luminosity range of the classical dSphs. 

Figure 3 shows the resulting $M_V$ and \vone\ for the VL2 satellites.  The
top panel shows the results directly from the VL2 catalog at $z=0$, while the 
bottom panel shows the results after the destruction, heating, and velocity 
corrections considered in this paper.  Filled red data points are those subhalos 
that are likely to survive tidal effects and retain enough baryons to be luminous, 
circles with `x' through them are those that are likely to have been destroyed 
by tidal effects in the presence of baryons, and empty circles are those that 
fall below the characteristic mass to retain baryons and form stars (see Figure 
2).  The filled black circles in Figure 3 identify a population of satellites 
that have lost more than 90\% of their mass since infall, but do not meet the 
destruction criteria that we outlined above.  \citet{Penarrubia2008} found 
that halos that lose more than 90\% of their mass begin to strip stars (see 
also Z12).  Hence, the luminosities of the black points in Figure 3 should 
be considered upper limits, as these halos will likely have had stars stripped 
and be fainter than at infall.  

Most of the brightest ($M_V < -12$) subhalos meet our criteria for destruction.  
We verified that, with the exception of one massive satellite accreted at 
$z=0.5$, all of these luminous, destroyed satellites were 
accreted $z > 1.5$.  The majority have 
$\vpeak > 40$ \kms, and by definition have orbits that take them within 20 kpc 
of the center of the parent halo.  This is consistent with the idea that early, 
massive satellites contribute to the growth of the inner stellar halo in 
Milky Way-mass galaxies \citep{Bullock2005, Zolotov2009}.  In DM-only runs that 
neglect the presence of the disk, some of the cuspy inner remnants of these early, 
massive satellites are capable of surviving to $z=0$.

We note that $M_V = -7$ corresponds to $\vinf = 13.3 \kms$.  The subhalos 
fainter than $M_V = -7$ in Figure 3  are those that had $\vpeak > 13.3$ \kms\ but 
a $\vinf < 13.3 \kms$.  Had we adopted \vpeak\ to assign stellar masses, all 
subhalos in Figure 3 would be brighter than $M_V = -7$.  Thus, the luminosities 
below $M_V \sim -9$ should be taken with caution, but we verfied that using \vpeak\ 
instead of \vinf\ had almost no impact on the luminosities of the satellites 
brighter than $M_V = -9$.  Hence, whether we adopt \vpeak\ or \vinf\ has no effect 
on the number, masses, or luminosities of surviving, luminous subhalos brighter 
than $M_V \sim -9$.  This luminosity corresponds to the faintest of the 
classical MW dwarf 
satellites.  Hence, use of \vinf\ does not change our conclusions regarding 
whether baryonic physics can address the massive missing satellites problem.   

Before applying the Z12 correction, the VL2 run contains more than 20 luminous 
satellites that have $\vone\ > 20$ \kms, completely inconsistent with the 
satellite population of either the MW or M31.  After 
applying the Z12 correction and considering satellites that are likely to be 
destroyed by baryonic physics or remain dark, only 3 satellites with 
$\vone >$ 20 \kms\ remain.  These 3 satellites are all more luminous than 
Fornax, the MW's brightest dSph, which we discuss further in the next section.

It can be seen from Figure 3 that applying the Z12 correction over-corrects 
a small number of subhalos, resulting in negative velocities.  These are a 
population of halos that have lost so much mass after infall that they have very 
low \vone\ values at $z=0$, and result in an overcorrection.  While Figure 3 
already identifies these as halos that have lost at least 97\% of 
their mass since infall, we verified that they actually lost more than 99.9\%
of their mass after infall, and that the majority have tidal radii less than 
1 kpc, and thus can safely be associated with destroyed subhalos.

\section{Discussion}

Two conclusions can be drawn from Figure 3.  The first conclusion is that 
a correction such as that suggested by Z12 is required to bring the masses 
(and hence velocities) of the predicted subhalo population in line with 
observational results.  Neglecting the baryonic effects that reduce the central 
masses of subhalos will inevitably lead to a population of subhalos substantially 
more massive (\vone\ $>$ 20 \kms) than any of the dSphs observed around the 
Milky Way.  The second conclusion is that both tidal destruction in 
the presence of a baryonic disk and UV heating must be considered in order 
to bring the total number of luminous satellites in line with observations.
All subhalos with $M_V$ brighter than $-10$ should be bright enough to be 
detected around our MW.  The uncorrected VL2 catalog would suggest that more 
than 100 detectable subhalos should exist, grossly inconsistent with the 
classical dSph population (roughly a dozen in the MW and two dozen in M31).
Considering destruction mechanisms alone does not bring the observable 
number into line with observations.  An additional correction that assumes 
suppression of star formation is necessary to further reduce the number of 
luminous satellites. 

Considering the effects of tidal destruction and heating can substantially 
reduce the number of luminous satellites that are predicted to survive 
at $z=0$ around a Milky Way-mass galaxy.  The exact number depends on the 
assumptions adopted, particularly for the number of dark satellites. The 
model adopted in this work is not intended to be conclusive, but rather to 
motivate more rigorous work on this topic.  Ideally, future work will adopt 
a semi-analytic model that follows the growth and merger history of individual 
halos to determine if they are massive enough to retain baryons and form 
stars.  More work is also needed to understand the influences of the disk 
on the survival of substructure.  Previous work has examined the influence 
of tidal stripping, though not always with the added presence of a galaxy 
disk \citep{Taylor2001, Stoehr2002, Hayashi2003, Kravtsov2004, Kazantzidis2004, 
Read2006, Penarrubia2008, penarrubia, Romano-Diaz2010, Nickerson2011}.  In this 
work, we have emphasized the effect of tidal stripping on satellites in the 
presence of a host with a baryonic disk \citep[or equivalently, any 
concentration of centralized baryons, e.g., at high $z$ the central baryons 
may not be a fully stable disk, but the concentration will still tidally 
impact the substructure, see][]{Chang2012}.  The Z12 correction, however, neglects disk 
shocking that occurs when a subhalo passes directly 
through the baryonic disk \citep{Taylor2001, D'Onghia2010}, and will lead 
to even faster disruption of a subhalo.\footnote{Z12 did have subhalos that 
experienced disk shocking, but these few halos were disrupted so strongly 
compared to other subhalos with larger pericenterric distances that 
the correction considers them outliers.}  Tidal heating of subhalos should 
also occur \citep{Gnedin1999, Mayer2001, D'Onghia2010b, Kazantzidis2011}, 
but requires very high resolution to capture \citep{Choi2009} and is unlikely 
to be accounted for properly in the Z12 correction.  Hence, all of these 
processes require more thorough study to fully understand the influence of 
baryons on the evolution of satellites.  Finally, one galaxy simulation alone 
is not sufficient to understand whether baryonic processes can solve the MSP.  
A statistical sample of MW-mass realizations must be used to quantify the 
theoretical predictions.  

By the same token, one observed galaxy alone is not 
sufficient to understand whether baryonic processes can solve the MSP.  
Looking at two galaxies, the MW and M31, it is already clear that the 
subhalo population of Milky Way-mass galaxies can vary significantly.
The surviving luminous VL2 satellites (shown as red circles in Figure 3) 
include 3 satellites with luminosities and velocities larger than the dSph 
population of the MW.  However, M31 does have such galaxies, and the 
surviving VL2 subhalos would more closely resemble the luminous satellite 
population of M31.  M31 has at least 4 luminous satellites with magnitudes 
brighter than $M_V < -14$.  Three of these are dwarf ellipticals, a 
population that is missing in the MW. 
All 3 of the VL2 subhalos were accreted at $z > 1$, have orbital pericenters 
under 50 kpc, and are unlikely to retain gas, making them probably look very 
similar to the dwarf ellipticals of M31.  We also note that many of the VL2 
satellites identified as luminous at $z=0$ reach \vpeak\ at relatively low $z$ 
(see Figure 2), and may have extended star formation to quite low $z$.  Z12 
also found that satellites with $M_V < -8$ had extended star formation 
histories, consistent with observational estimates \citep{Grebel2004, 
Dellenbusch2008, Weisz2011}.  We anticipate that the more recently 
discovered ultra-faint satellites should have had their star formation 
truncated at high $z$ \citep{Brown2012}.

In comparing our results for the luminous satellites of VL2 to those 
of the MW and M31, we should consider the impact of halo host mass on 
satellite populations. It has been shown that the number of subhalos at 
a given mass scales with host mass, since the mass assembly of DM halos 
is self-similar \citep{Stewart2008, Fakhouri2010}.  Several works have 
argued that the halo mass of M31 is nearly twice as massive as the 
MW's halo mass 
\citep[e.g.,][]{Kallivayalil2009, Guo2010, Watkins2010}, suggesting that 
M31 should have both a brighter satellite population, and more massive 
satellites,  than the MW. While the exact halo mass of the MW is still 
unknown, recent estimates \citep[e.g.,][]{Smith2007, Xue2008, Guo2010, 
oGnedin2010} find a halo mass range of $0.7 - 2.0 \times 10^{12} 
M_{\odot}$.  The virial mass of VL2 ($1.7 \times 10^{12} M_{\odot}$) is 
near the upper range of the MW's mass estimates, but perhaps much closer 
to the virial mass of M31 \citep{Watkins2010}, particularly if M31 is 
twice as massive as the MW. Therefore, the resemblance of the VL2 bright 
satellite population, i.e., its 3 satellites with $M_V < -14$, to the 
bright satellites of M31 is perhaps indicative that the halo mass of VL2 
more closely matches the halo mass of M31 than the MW 
\citep[see also][]{Wang2012, Vera-Ciro2012}.

Because this simulated population is a better match to M31 than the MW, 
we should also consider how common the MW satellite distribution is.  
A few authors have recently begun exploring this question.  On the 
observational side, searches for satellites in the SDSS around galaxies as 
luminous as the MW find $\sim$10 or fewer satellites more luminous than 
Fornax \citep{Busha2011, Guo2011, Lares2011, Strigari2012}.  Adopting a 
semi-analytic model to assign stellar masses, \citet{Vera-Ciro2012} find 
that 2 to 5 subhalos brighter than $M_V < -14$ consistently exist in the 
six MW-mass halos in the high resolution Aquarius DM-only runs.  Using a 
larger statistical sample, both \citet{Wang2012} and \citet{Purcell2012} 
find a 10-20\% probability of finding a subhalo population similar to the 
MW's at a halo mass of 10$^{12}$\Msun.  This suggests that the MW is 
somewhat rare, though not exceedingly so.  The rarity of the MW's subhalo 
luminosity function seems to be linked to a gap in luminosities between 
Fornax (at $M_V = -13.4$) and the SMC (at $M_V = -16.8$), though more 
work needs to be done to quantify how common the gap is, and if having 
a pair of galaxies as bright as the Magellanic Clouds has any influence 
of the existence of such a gap.

Despite the caveats listed above, it is clear that 
baryons offer a promising solution to solving the MSP, without an additional 
form of warm or self-interacting dark matter.  Certainly heating and destruction 
have long been considered potential solutions to bring the number of 
predicted satellites in line with the number of observed satellites 
\citep{Bullock2000, Somerville2002, Kravtsov2004, SatKinematics, Koposov2009, 
Nickerson2011, Penarrubia2012}.
The key point, however, is that even if the number of luminous satellites 
can be brought into line, the overall distribution in masses (and 
observed velocities) of those satellites is usually too high \citep{toobig, 
Boylan-kolchin2012}.  The Z12 model builds on this previous work by 
considering the additional impact of baryons on the dark matter density 
slopes of satellites, finding shallow profiles in satellites brighter 
than $M_V < -12$, which further reduces their mass.  When combined with 
more effective tides after infall \citep{Taylor2001, Stoehr2002, Read2006, 
penarrubia}, both the {\it number} and {\it masses} of the subhalo 
population can be brought into agreement with observations for the first 
time.

In fact, the model utilized by Z12 goes beyond solving the MSP or TBTF
problems.  The model in Z12 has also been shown to create bulgeless dwarf 
disk galaxies \citep{Governato2010} and smaller bulges in higher mass 
galaxies \citep{Brook2012}.  This is because winds driven by supernovae 
preferentially remove low angular momentum material from galaxies 
\citep[Christensen et al.\,2012;][]{Brook2011}.  Importantly, these winds 
are naturally driven (rather than artificially inserted) when supernovae 
deposit thermal energy into a high density medium, and create highly 
overpressurized bubbles.  Hence, a model must be able to reproduce the 
high densities found in star forming molecular clouds ($> 10-100$ atoms/cc) 
to naturally drive such winds.  The ability to model such 
high density regions is what allowed for dark matter core creation in the 
dSph satellites simulated in Z12.  Previous studies of dSphs in MW-mass 
satellites allowed for star formation at lower densities ($<$ 1 amu/cc), 
and hence found that baryons either steepened the inner densities or 
had little affect \citep{Sales2007, Okamoto2010, Nickerson2011, Wadepuhl2011, 
Parry2012, DiCintio2012, Sawala2012}.  Critically, the
expanding overpressurized bubbles are able to drive rapid fluctuations 
in the potential well of a galaxy.  The cumulative effect is to expand 
the central orbits of dark matter, 
transforming an initially steep, cuspy DM density profile into a shallow, 
cored profile \citep[e.g.,][]{Read2005, Mashchenko2006, Pontzen2012, Maccio2012, 
Pontzen2012, Teyssier2012}.  \citet{Governato2012} and \citet{Oh2011} showed 
that the density profiles of simulated galaxies within this model are in 
excellent agreement with the high resolution observations of field galaxies.  

In summary, the model adopted by Z12 not only yields a population of 
satellite galaxies in agreement with MW or M31 observations, potentially 
solving both the ``missing satellites'' and ``too big to fail'' problems, 
but it also simultaneously addresses other small scale problems within 
CDM.  Hence, a unified baryonic solution 
remains viable to solve the small scale crisis of CDM.  Future work 
must quantify the impact of baryons on the DM structure of galaxies, and 
make testable predictions across all galaxy masses.

\section{Conclusions}

It has long been appreciated that the observed population of Milky Way 
satellites is at odds with the distribution predicted by dark matter-only 
simulations. In this article, we have argued that effects associated with 
baryonic physics can reconcile the results of such simulations with observations, 
without the need for dark matter that is warm, self-interacting, or with 
properties that are otherwise different from those of the standard cold and 
collisionsless paradigm. In particular, supernova feedback in luminous satellites 
can reduce the density of dark matter in the inner volumes of these systems, 
while the presence of a baryonic disk can enhance the degree of tidal stripping 
that takes place. The combination of these effects leads to a reduction of the 
masses of the predicted subhalo population, bringing the overall number of 
satellites into concordance with that observed. In particular, in the Via 
Lactea II sample considered, we found that these effects reduced the number of 
subhalos with masses larger than seen in the MW satellites 
from more than 20 to only a few. 

After determining the distribution of massive dark matter subhalos predicted to be present in a Milky Way-mass galaxy, we turned our attention on the question of how many of these objects are likely to host luminous satellites. For reasonable assumptions regarding the stellar-to-halo mass relationship and for the criteria for destruction via tidal stripping, we predict a luminous satellites population that is in adequate agreement with both the Milky Way and M31.

While the work presented here is not intended to represent the final word on this topic, we have shown that baryonic effects can lead to a population of satellites around Milky Way-mass galaxies that is is good agreement with observations. This strongly reduces the motivation for warm or self-interacting dark matter scenarios and represents yet another success for cold dark matter.

\acknowledgments
AB acknowledges support from The Grainger Foundation.  AZ acknowledges 
support from the Lady Davis Foundation.  AZ's work was partially supported 
by the ISF grant 6/08, by GIF grant G-1052-104.7/2009, by the DFG  grant 
STE1869/1-1.GE625/15-1.  This work was supported in part by the U.S. 
National Science Foundation, grants OIA-1124453 (PI P.~Madau) and 
OIA-1124403 (PI A.~Szalay).

\end{document}